\documentclass[twocolumn,showpacs,aps,pra,amsfonts,amsmath,amssymb,floatfix]
{revtex4}
\usepackage[dvips]{graphicx}

\newcommand{\affA}{%
     National Institute of Information and Communications Technology, \\
     4-2-1 Nukui-kitamachi, Koganei, Tokyo 184-8795, Japan \\
     CREST, Japan Science and Technology Agency, 
     1-9-9 Yaesu, Chuoh-ku, Tokyo 103-0028, Japan}

\begin{document}
\title{Conditional generation of an arbitrary superposition of 
coherent states
}
\date{\today}
\author{Masahiro Takeoka}
\author{Masahide Sasaki}
\address{\affA}%

\begin{abstract}

We present a scheme to conditionally generate an arbitrary superposition of 
a pair of coherent states from a squeezed vacuum by means of 
the modified photon subtraction where  
a coherent state ancilla and two on/off type detectors are used. 
We show that, 
even including realistic imperfections of the detectors, 
our scheme can generate a target state with a high fidelity. 
The amplitude of the generated states can be amplified by 
conditional homodyne detections. 

\end{abstract}

\maketitle

\section{Introduction
\label{sec:intro}}
Conditional quantum operation based on photon detection 
plays an important role 
in recent optical quantum information processing. 
Particularly, in the continuous variable regime, 
it is the only available tool with current technology 
to generate non-Gaussian states from Gaussian ones. 
A typical example is the `photon subtraction' 
operation, in which a nonclassical input state is split 
by a highly transmissive beamsplitter (BS) and 
the reflected state is measured by a photon number resolving detector (PNRD). 
Selecting the event that the detector observes photons, 
one obtains a non-Gaussian transformation
from the input to output quantum state.

This type of conditional operation was formulated in \cite{Ban94}.  
Dakna {\it et al.} \cite{Dakna97} then showed 
that applying it to a squeezed vacuum, one can generate 
a non-Gaussian state which is close to 
the superposition of coherent states with plus or minus phase 
\begin{equation}
\label{eq:cat_state}
|C_\pm(\alpha)\rangle = \frac{1}{\sqrt{\mathcal{N}_\pm}} 
\left( |\alpha\rangle \pm |-\alpha\rangle \right) ,
\end{equation}
with very high fidelity, 
where $|\alpha\rangle$ is a coherent state with the amplitude of $\alpha$ 
and $\mathcal{N}_\pm$ is the normalization factor.

Recently, such state has been experimentally generated 
by a single photon subtraction from 
pulsed \cite{Wenger04,Ourjoumtsev06} 
and CW \cite{N-Nielsen06,Wakui06} squeezed vacua. 
In these experiments, since a reflected beam includes sufficiently 
small average number of photons ($\bar{n} \ll 1$), 
single photon detection was approximately 
realized by use of an avalanche photodiode (APD) 
which is often called as `on/off' type detector 
since it discriminates only a presence of photons 
instead of resolving photon numbers.

The progress of these experiments promises the realizations of 
more complicated applications of photon subtraction proposed so far, 
including the improvement of quantum teleportation 
\cite{Opatrny00,Cochrane02,Olivares03} 
and entanglement-assisted coding \cite{Kitagawa05}, 
entanglement distillation \cite{Browne03}, 
loophole free tests of Bell's inequalities 
\cite{Nha04,G-Patron04,G-Patron05}, 
and optical quantum computations 
in quadrature basis \cite{Gottesman01,Menicucci06} or 
superposed coherent state basis 
\cite{Ralph03,Lund05}. 
In the last application, 
$|C_\pm(\alpha)\rangle$ with appropriate $\alpha$ is 
required as an ancillary state. 
To prepare such ancillae, the method to conditionally amplify 
$\alpha$ with on/off detectors has been 
proposed \cite{Lund04,Jeong05}. 
Fiur\'{a}\v{s}ek {\it et al.} \cite{Fiurasek05} also recently showed that 
one can arbitrarily synthesize 
a single-mode quantum state up to the $N$-photon eigenstate 
by concatenating squeezing operations and $N$ times single photon 
subtractions \cite{Fiurasek05}.

In this paper, 
we propose a method to conditionally generate the state in which 
two coherent states are superposed with {\it arbitrary} ratio and phase, 
$c_+ |\alpha\rangle + c_- |-\alpha\rangle$. 
This is accomplished by a simple modification of the scheme 
proposed by Dakna {\it et al.} (DAOKW) \cite{Dakna97}. 
We first discuss an ideal scheme using two PNRDs 
and a qubit ancilla, which 
produces a superposition of the one- and two-photon subtracted 
states.  
We show that such state 
fairly well approximates the target state 
$c_+ |\alpha\rangle + c_- |-\alpha\rangle$. 
We next present a more practical scheme 
where PNRDs and a qubit ancilla are replaced by 
the on/off detectors and a coherent state ancilla. 
Even including practical imperfections of the detectors, 
it can generate the target state with a high fidelity.

Our scheme should be compared with the one 
by Fiur\'{a}\v{s}ek {\it et al.} \cite{Fiurasek05}, 
which requires $N$ detectors to synthesize a state consisting of 
the number states up to $|N\rangle$. 
Ours, on the other hand, uses only two detectors to synthesize 
a fully continuous variable state while, in return, 
the class of states to be generated is restricted.

Finally, we show that our scheme is useful to simplify 
the setup of the conditional amplification of the superpositions of 
coherent states 
originally proposed in \cite{Lund04,Jeong05}.

The paper is organized as follows. 
In Sec.~\ref{sec:PNRDscheme}, we discuss an ideal setup 
with PNRDs and a qubit ancilla and how the scheme's parameters 
are optimized to generate desired superposed states. 
In Sec.~\ref{sec:on/offscheme}, a practical scheme using on/off detectors 
and a coherent state ancilla is shown and its experimental feasibility 
is numerically examined. 
An application of our scheme to the conditional amplification 
of superposed coherent states is shown in Sec.~\ref{sec:amp} 
and Sec.~\ref{sec:conclusion} concludes the paper.

\section{Generation of an arbitrarily superposition of coherent states
\label{sec:PNRDscheme}}
\begin{figure}
\begin{center}
\includegraphics[width=1\linewidth]{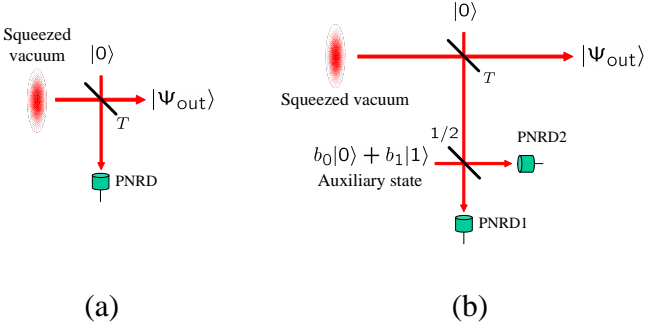}   %
\caption{\label{fig:SchemePNRD} 
(a) Generation of a plus- or minus-superposition of coherent states 
via photon subtraction operation with a photon number resolving detector 
(PNRD). 
(b) Generation of an arbitrary superposition of coherent states 
with PNRDs and a qubit ancilla. 
}
\end{center}
\end{figure}

Figure~\ref{fig:SchemePNRD}(a) illustrates 
the DAOKW photon subtraction scheme \cite{Dakna97}. 
A squeezed vacuum with the squeezing parameter $r$ 
is mixed with a vacuum by a highly transmissive BS
and the reflected part of the state is detected by a PNRD. 
When the reflected part is projected onto the photon number 
eigenstate $|m\rangle$ ($m>0$), 
the state remained in the transmitted mode is 
reduced to be the $m$ photon subtracted squeezed vacuum state 
that can be described by a minus- or plus-superposition of 
two distinct states as
\begin{equation}
\label{eq:Dakna_decomposition}
|\Psi_m\rangle = A (|\Psi_m^{(+)}\rangle + (-1)^m |\Psi_m^{(-)}\rangle) , 
\end{equation}
where $A$ is the normalization factor \cite{Comment1}. 
It was shown that, 
with an appropriate input squeezed vacuum, 
the states $|\Psi_m^{(\pm)}\rangle$ are very close to 
the coherent states $|\pm\alpha_m\rangle$ 
and thus the states $|\Psi_m\rangle$ are 
also very close to a superposition of coherent states.

Let us extend the above scheme as illustrated 
in Fig.~\ref{fig:SchemePNRD}(b). 
Let the upper BS has the power transmittance $T \approx 1$ and 
the lower be a balanced BS. 
The reflected part of the state is mixed with the auxiliary state 
$b_0 |0\rangle + b_1 |1\rangle$ and then each port is incident 
into a PNRD. 
After some calculations, one finds that if the measurement outcome 
of the two detectors is (2, 0) or (0, 2), the reflected part 
is effectively projected onto 
$\mp b_1^*/\sqrt{2} |1\rangle + b_0^*/\sqrt{2} |2\rangle$ 
and the transmitted state conditioned on either of these outcomes 
has the form 
\begin{equation}
\label{eq:proj_ideal}
|\Psi_{\rm out}\rangle = a_1 |\Psi_1\rangle + a_2 |\Psi_2\rangle , 
\end{equation}
where $a_1$ and $a_2$ are the functions of $b_0$, $b_1$, and $T$. 
Since $|\Psi_m\rangle$ can be regarded as a superposition specified in 
Eq.~(\ref{eq:cat_state}), the state $|\Psi_{\rm out}\rangle$ is also 
expected to be a superpostion of $|\pm\alpha\rangle$ with 
the controlled ratio and phase 
by choosing ancilla parameters $b_0$ and $b_1$ appropriately.

\begin{figure}
\begin{center}
\includegraphics[width=0.8\linewidth]{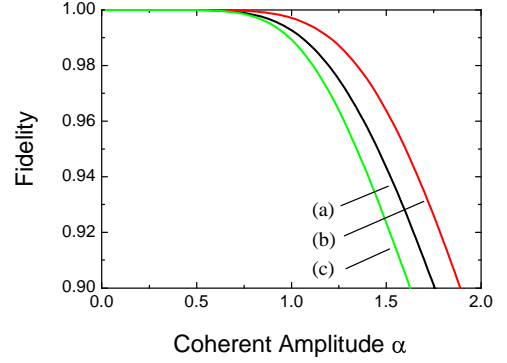}   %
\caption{\label{fig:Fidelity}
Fidelities between the photon subtracted states and 
the ideal superposition of coherent states. 
(a) $|\langle \alpha|\phi_+ \rangle|^2$, 
(b) $|\langle C_-(\alpha)|\Psi_1\rangle|^2$, and 
(c) $|\langle C_+(\alpha)|\Psi_2\rangle|^2$. 
}
\end{center}
\end{figure}

Now let us see the state in Eq.~(\ref{eq:proj_ideal}) more carefully.  
To show $|\Psi_{\rm out}\rangle$ to be a superposition of 
two (classical) macroscopically distinct states, 
one has to find the following decompositions, 
\begin{eqnarray}
\label{eq:decompose1}
|\Psi_1\rangle & = & \frac{1}{2c_1} 
\left( |\phi_+\rangle - |\phi_-\rangle \right) , 
\\
\label{eq:decompose2}
|\Psi_2\rangle & = & \frac{1}{2c_2} 
\left( |\phi_+\rangle + |\phi_-\rangle \right) , 
\end{eqnarray}
in which $|\phi_{\pm}\rangle$ are close enough to the coherent states 
$|\pm\alpha\rangle$. $c_1$ and $c_2$ are the normalization factors 
satisfying $|\phi_\pm\rangle = c_2 |\Psi_2\rangle \pm c_1 |\Psi_1\rangle$. 
Note that the decomposition described in Eq.~(\ref{eq:Dakna_decomposition}) 
\cite{Dakna97} is not optimal in our purpose 
since $|\Psi_1\rangle$ and $|\Psi_2\rangle$ 
do not share the common decomposed components.

The optimal $|\phi_\pm\rangle$ to maximize the fidelity 
$|\langle\phi_\pm|\pm\alpha\rangle|^2$ can be 
derived from the exact expression of $|\Psi_m\rangle$ \cite{Comment1} 
and Eqs.~(\ref{eq:decompose1}) and (\ref{eq:decompose2}) 
as 
\begin{eqnarray}
\label{eq:c1}
c_1 & = & \sqrt{\frac{3\lambda T}{(1+\lambda T)(1+2\lambda T)}} , \\
\label{eq:c2}
c_2 & = & \sqrt{\frac{1+2\lambda^2 T^2}{(1+\lambda T)(1+2\lambda T)}} ,
\end{eqnarray}
where $\lambda = \tanh r$ is the squeezing parameter and 
the amplitude of the corresponding coherent states is given by 
\begin{equation}
\label{eq:coherent_state}
|\pm\alpha\rangle = 
\left| \pm\sqrt{\frac{3\lambda T}{1-\lambda^2 T^2}} \right\rangle .
\end{equation}
Then we have quasi-coherent states 
\begin{eqnarray}
\label{eq:phi_pm}
|\phi_\pm\rangle & = & 
\frac{(1-\lambda^2 T^2)^{3/4}}{2\sqrt{(1+\lambda T)(1+ 2\lambda T)}} 
  \sum_{n=0}^\infty \frac{(2n+2)!}{(n+1)!} \left(\frac{\lambda T}{2}\right)^n 
\nonumber\\  & & 
\left( \frac{1-\lambda^2 T^2}{\sqrt{(2n)!}} |2n\rangle 
  \pm \sqrt{\frac{3\lambda T}{(2n+1)!}} |2n+1\rangle \right), 
\nonumber\\
\end{eqnarray}
and the fidelity between Eqs.~(\ref{eq:coherent_state}) and 
(\ref{eq:phi_pm}) is given by 
\begin{eqnarray}
\label{eq:fidelity}
F & = & |\langle\alpha|\phi_+\rangle|^2
\nonumber\\ & = & 
\sqrt{1- \lambda^2 T^2} (1+\lambda T)(1+2\lambda T) 
\exp\left[ -\frac{3\lambda T}{1+\lambda T} \right] ,
\nonumber\\
\end{eqnarray}
which is plotted in Fig.~(\ref{fig:Fidelity}) by the black line (line (a)). 
For $\alpha < 1$, more than $0.99$ fidelity is achieved. 
The validity of this optimization is also confirmed by 
looking at the fidelities $|\langle\Psi_1|C_-(\alpha)\rangle|^2$ and 
$|\langle\Psi_2|C_+(\alpha)\rangle|^2$ for the same $\alpha$. 
These are plotted in the same figure by the red (line (b)) and 
green (line (c)) lines, respectively.

\section{Practical setup with on/off detectors
\label{sec:on/offscheme}}
Preparing PNRDs and a photon number qubit ancilla 
is still somehow challenging with current technology. 
In this section, we show a modified and more practical scheme in which 
PNRDs and a qubit ancillary state are replaced 
with on/off detectors and a coherent state, respectively.

\begin{figure}
\begin{center}
\includegraphics[width=0.8\linewidth]{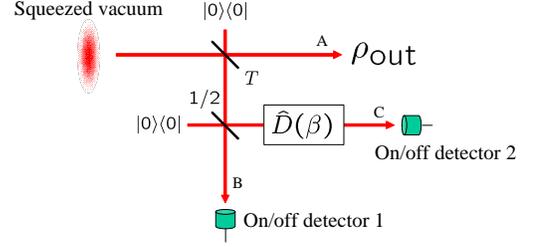}   %
\caption{\label{fig:SchemeOnOff}
A schematic of the generation of an aribitrary superposition of 
coherent states 
with on/off detectors and a displacement operation. 
}
\end{center}
\end{figure}

The modified scheme is depicted in Fig.~\ref{fig:SchemeOnOff}. 
The reflected state from the first BS with the transmittance $T$ 
is further split by the second balanced BS. 
One beam is directly measured by an on/off detector (mode B) 
and the other is first shifted by the displacement operator 
$\hat{D}(\beta) = \exp[ \beta \hat{a}^\dagger - \beta^* \hat{a} ]$ and 
then measured by another on/off detector (mode C). 
It is well known that a displacement operation is 
realized by interfering the signal with 
an auxiliary coherent state $|\beta/\sqrt{1-T_D}\rangle$ 
by a BS with the transmittance $T_D$. 
In the limit of $T_D \to 1$, this operation is exactly the 
same as $\hat{D}(\beta)$. 
The output state is conditionally selected only when 
both detectors are simultaneously clicked by photons. 
The photons detected in mode B always come from the squeezed vacuum 
while, in mode C, the photons from the squeezed vacuum are interfered 
by the displacement. 
This quantum interference and the on/off detection 
realizes a projection 
onto a superposition of different photon number states.

The positive operator-valued measure (POVM) for 
on/off detectors 
is described by $\{ \hat{\Pi}_{\rm off}, \hat{\Pi}_{\rm on} \}$ 
where $\hat{\Pi}_{\rm off} = |0\rangle\langle0|$ and 
$\hat{\Pi}_{\rm on} = \hat{I} - \hat{\Pi}_{\rm off}$ and 
$\hat{I}$ is an identity operator. 
Similarly, when a displacement operation $\hat{D}(\beta)$ is placed 
before the detector, as in mode C, 
the total on/off POVM is expressed as 
\begin{eqnarray}
\label{eq:off_POVM_with_displacement}
\hat{\Pi}_{\rm off} (\beta) 
& = & \hat{D}^\dagger(\beta)|0\rangle\langle0| \hat{D}(\beta) 
= |-\beta\rangle\langle-\beta|, 
\\
\label{eq:on_POVM_with_displacement}
\hat{\Pi}_{\rm on} (\beta) 
& = & \hat{D}^\dagger(\beta) (\hat{I} - |0\rangle\langle0|) \hat{D}(\beta) 
\nonumber\\ 
& = & \hat{I} - |-\beta\rangle\langle-\beta| .
\end{eqnarray}
The average photon number reflected to mode B 
from the initial squeezed vacuum 
is given by $(1-T)\sinh^2 r$. 
For moderate squeezing, 
this is sufficiently small
to assume that the reflected beam in mode B 
contains maximally one photon and to ignore the more than 
one photon eigenspace at the measurement process 
(e.g. $(1-T)\sinh^2 r \sim 0.005$ for $r=0.3$ and $T=0.95$). 
When $|\beta|^2$ in Eqs.~(\ref{eq:off_POVM_with_displacement}) 
and (\ref{eq:on_POVM_with_displacement}) 
is also sufficiently small such that one can approximate as
$|-\beta\rangle \approx |0\rangle - \beta|1\rangle$, 
$\hat{\Pi}_{\rm off} (\beta)$ in mode C acts as 
the projection onto $|0\rangle - \beta|1\rangle$, 
and $\hat{\Pi}_{\rm on} (\beta)$ as 
the projection onto the orthogonal superposition 
$\beta^* |0\rangle + |1\rangle$.
As a consequence, when both detectors are clicked, 
the reflected part of the state is projected onto 
\begin{eqnarray}
\label{eq:measurement_vector}
&&
_C \langle 0| \hat{B}^\dagger_{1/2} 
|1\rangle_B \left( \beta^* |0\rangle_C + |1\rangle_C \right) 
\nonumber\\ && \propto \, 
_C \langle 0| \left\{ -\beta^* (|01\rangle - |10\rangle) 
- (|02\rangle - |20\rangle) \right\}_{BC}
\nonumber\\ && = 
\beta^* |1\rangle_B + |2\rangle_B , 
\end{eqnarray}
where the normalization factors and global phases are omitted and 
$\hat{B}_T = \exp[\theta (\hat{a}^\dagger \hat{b} - \hat{a} \hat{b}^\dagger)]$ 
is a BS operator with $\cos\theta=\sqrt{T}$.

On the other hand, the state after the first BS is described as 
\begin{eqnarray}
\label{eq:BS1}
\hat{B}_T \left( \hat{S}(r)|0\rangle_A \right) |0\rangle_B 
& = & \sqrt{\mathcal{P}_0} |\Psi_0\rangle_A |0\rangle_B 
+ \sqrt{\mathcal{P}_1} |\Psi_1\rangle_A |1\rangle_B 
\nonumber\\ && 
+ \sqrt{\mathcal{P}_2} |\Psi_2\rangle_A |2\rangle_B 
+ \cdots
\end{eqnarray}
where $S(r) = \exp[ r/2 (\hat{a}^2-\hat{a}^{\dagger\,2})]$ 
is a squeezing operator and 
$\mathcal{P}_m$ is the probability to observe $m$ photons 
in modes $B$ and $C$ \cite{Dakna97}, 
\begin{eqnarray}
\label{eq:P_m}
\mathcal{P}_m & = & 
\sqrt{\frac{1-\lambda^2}{1-\lambda^2 T^2}} 
\left[ \frac{\lambda^2 T^2 (1-T)}{T(1-\lambda^2 T^2)} \right]^m 
\nonumber\\ && 
\times \sum_{k=0}^{[m/2]} \frac{m!}{(m-2k)!(k!)^2(2\lambda T)^{2k}} .
\end{eqnarray}
From Eqs.~(\ref{eq:measurement_vector}) and (\ref{eq:BS1}), 
we have the conditional output for the simultaneous click as 
\begin{equation}
\label{eq:cond_output}
|\Psi_{\rm out}\rangle \propto 
\beta \sqrt{\mathcal{P}_1} |\Psi_1\rangle 
+ \sqrt{\mathcal{P}_2} |\Psi_2\rangle .
\end{equation}
Consequently, for the generation of the superposition state 
$c_+ |\phi_+ \rangle + c_- |\phi_-\rangle$, 
the optimal displacement $\beta$ is derived from 
Eqs.~(\ref{eq:decompose1}$-$\ref{eq:c2}),
(\ref{eq:P_m}), and (\ref{eq:cond_output}) and given as 
\begin{equation}
\label{eq:displacement_beta}
\beta = \frac{c_+ - c_-}{c_+ + c_-} 
\left( \frac{3\lambda (1-T)}{2 (1-\lambda^2 T^2)} \right)^{1/2} ,
\end{equation}
which is valid under the condition of $|\beta|^2 \ll 1$, i.e.  
\begin{equation}
\label{eq:beta_condition}
\left| \frac{c_+ - c_-}{c_+ + c_-} \right|^2 \ll 
\frac{2 (1-\lambda^2 T^2)}{3\lambda (1-T)} .
\end{equation}
Note that Eq.~(\ref{eq:displacement_beta}) 
is almost optimal for arbitrary $\beta$ 
although this condition will be broken when $c_+ + c_- \sim 0$ 
i.e. one wants to generate $|\Psi_{\rm out}\rangle \sim |\Psi_1\rangle$. 
For large $|\beta|^2$, 
Eq.~(\ref{eq:on_POVM_with_displacement}) 
up to one photon state is given by 
\begin{equation}
\label{eq:beta_large}
\hat{\Pi}_{\rm on} (\beta) \to 
(1-e^{-|\beta|^2}) \hat{I} 
+ e^{-|\beta|^2} (\beta^* |0\rangle + |1\rangle)
(\beta \langle0| + \langle1|) . 
\end{equation}
Although it makes the output as a mixed state of 
\begin{equation}
\label{eq:mixed_out}
\hat{\rho}_{\rm out} =
(1-e^{-|\beta|^2}) |\Psi_1\rangle\langle\Psi_1| + 
e^{-|\beta|^2} |\Psi_{\rm out}\rangle\langle\Psi_{\rm out}| , 
\end{equation}
this is clearly a negligible error since 
$|\langle\Psi_{\rm out}|\Psi_1\rangle|^2$ exponentially 
approaches to unit.

In the rest of this section, we numerically examine 
the conditional outputs under realistic conditions. 
In practice, there is always finite probability to 
detect more than one photon at each detector. 
Moreover, the detectors themselves have finite imperfections. 
The POVM for an imperfect on/off detector 
with the displacement operation $\hat{D}(\alpha)$ 
is given by 
\begin{eqnarray}
\label{eq:on/off_POVM_off}
\hat{\Pi}_{\rm off}(\alpha,\eta,\nu) & = & 
e^{-\nu} \sum_{m=0}^\infty (1-\eta)^m 
\hat{D}^\dagger(\alpha) |m \rangle\langle m| \hat{D}(\alpha) , 
\nonumber\\
\\
\label{eq:on/off_POVM_on}
\hat{\Pi}_{\rm on}(\alpha,\eta,\nu) 
& = & \hat{I} - \hat{\Pi}_{\rm off}(\alpha) ,
\end{eqnarray}
where $\eta$ and $\nu$ are the quantum efficiency and dark count 
of the detector, respectively.

\begin{figure}
\begin{center}
\includegraphics[width=1\linewidth]{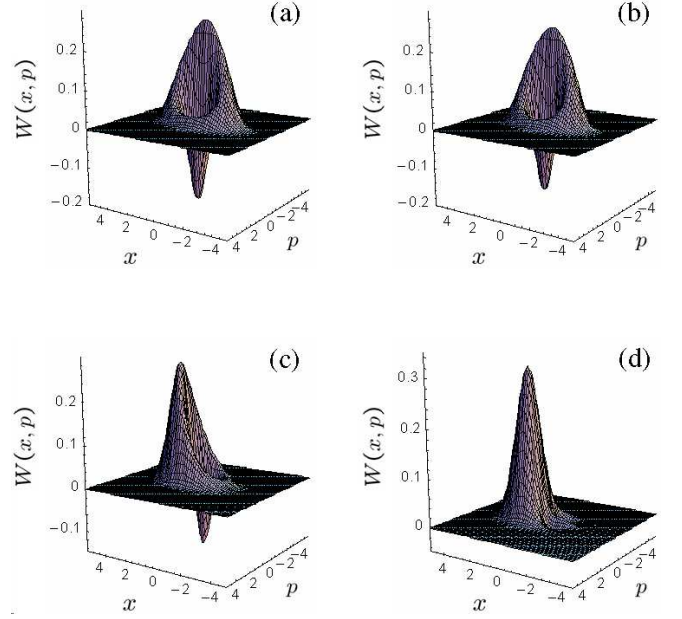}   %
\caption{\label{fig:WignerFunctions}
The Wigner functions of the superpositions of coherent states 
$( c_+ |\phi_+\rangle + c_- |\phi_-\rangle )/\mathcal{N}$ 
generated by the on/off detector photon subtractions illustrated 
in Fig.~(\ref{fig:SchemeOnOff}) 
for (a), (b) $\{c_+, c_-\} = \{1,i\}$, (c) $\{c_+, c_-\} = \{3,-1\}$, and 
(d) $\{c_+, c_-\} = \{1,0\}$ where 
(a) $r=0.3$, $T=0.999$, $\alpha=0.97$ with the ideal detectors of 
$\eta=1$, and $\nu=0$, and (b), (c), (d) $r=0.3$, $T=0.95$, $\alpha=0.95$, 
with the practical imperfection of detectors, $\eta=0.1$, and $\nu=10^{-7}$. 
The fidelities between the plotted state and 
the ideal state 
$(c_+ |\alpha\rangle + c_- |-\alpha\rangle)/\mathcal{N}$ 
are (a) $F=0.993$, (b) $F=0.952$, (c) $F=0.978$, and (d) $F=0.994$. 
}
\end{center}
\end{figure}

This kind of detectors makes the output an unwanted mixed state. 
To derive photon subtracted states under these conditions, 
it is useful to use the characteristic functions to describe 
the states and POVMs \cite{Kim05,Olivares05,Molmer06}. 
Since the input squeezed vacuum is a Gaussian state, 
its characteristic function can be described as 
\begin{equation}
\label{eq:CFSV}
\chi_{\rm SV} ({\bf \omega}) = \exp\left[ -\frac{1}{4} 
{\bf \omega}^T \Gamma_{\rm SV} {\bf \omega} \right] , 
\end{equation}
where ${\bf \omega} = (u, v)^T$ is a two dimensional vector and 
$\Gamma_{\rm SV}$ is the covariance matrix for the squeezed vacuum 
\begin{equation}
\label{eq:CovMxSV}
\Gamma_{\rm SV} = \left[
\begin{array}{cc}
e^{2r} & 0 \\
0 & e^{-2r} 
\end{array}
\right] .
\end{equation}
The mixing of a squeezed vacuum and a vacuum by a BS is described by 
a linear transformation 
\begin{equation}
\label{eq:BS_transformation}
\Gamma_{\rm SV} \oplus \Gamma_{\rm vac} \to 
S_{BS}^T (T) \Gamma_{\rm SV} \oplus \Gamma_{\rm vac} S_{BS} (T) , 
\end{equation}
where $\Gamma_{\rm vac} = {\bf I}$ is the covariance matrix for 
the vacuum state and $S_{BS} (T)$ is the $4 \times 4$ matrix 
\begin{equation}
\label{eq:BS_matrix}
S_{BS} (T) = \left(
\begin{array}{cc}
\sqrt{T} \, {\bf I} & \sqrt{1-T} \, {\bf I} \\
-\sqrt{1-T} \, {\bf I} & \sqrt{T} \, {\bf I} 
\end{array}
\right) . 
\end{equation}
Then the covariance matrix after the two BSs in Fig.~\ref{fig:SchemeOnOff} 
is given by 
\begin{eqnarray}
\label{eq:SV_after_2BS}
\tilde{\Gamma} & = &
(I \oplus S_{BS}^T (1/2)) (S_{BS}^T (T) \oplus I) 
(\Gamma_{\rm SV} \oplus \Gamma_{\rm vac} \oplus \Gamma_{\rm vac}) 
\nonumber\\ && \times
(S_{BS} (T) \oplus I) (I \oplus S_{BS} (1/2)) .
\end{eqnarray}
Let the characteristic function corresponding to $\tilde{\Gamma}$ 
be $\chi_{\rm in} ({\bf \omega_A}, {\bf \omega_B}, {\bf \omega_C})$. 
The output characteristic function conditioned on the simultaneous click 
in both two detectors is then given by 
\begin{eqnarray}
\label{eq:cond_output_CF}
\chi_{\rm out} ({\bf \omega_A}) & = & 
\int_{-\infty}^\infty \int_{-\infty}^\infty 
{\rm d}{\bf \omega_B} {\rm d}{\bf \omega_C} \, 
 \chi_{\rm in} ({\bf \omega_A}, {\bf \omega_B}, {\bf \omega_C})
\nonumber\\ & & \times
 \chi_{\rm on} (-{\bf \omega_B}, -{\bf \omega_C}) , 
\end{eqnarray}
where 
\begin{equation}
\label{eq:CF_for_POVM}
\chi_{\rm on} ({\bf \omega_B}, {\bf \omega_C}) 
= \chi_{\rm on } (0, {\bf \omega_B}) 
\chi_{\rm on } (-\beta^*, {\bf \omega_C}) ,
\end{equation}
and $\chi_{\rm on } (\alpha, {\bf \omega})$ corresponds to 
the characteristic function for the POVM defined 
in Eq.~(\ref{eq:on/off_POVM_on}). 
Finally, the Wigner function for the output is given 
by the Fourier transform of the characteristic function as 
\begin{equation}
\label{eq:cond_output_Wigner}
W_{\rm out} ({\bf z}) = \frac{1}{(2\pi)^2} 
\int_{-\infty}^\infty d{\bf \omega} \, \chi_{\rm out} ({\bf \omega}) 
\exp\left[-i{\bf \omega}^T {\bf z}\right] ,
\end{equation}
where ${\bf z} = ( x,p )^T$.

Figure~\ref{fig:WignerFunctions}(a) shows $W_{\rm out} ({\bf z})$ 
corresponding to the state $|\phi_+\rangle + i |\phi_-\rangle$ 
with nearly ideal parameters, $T=0.999$, unit quantum efficiency, 
and zero dark counts. 
The squeezing parameter $r=0.3$ corresponds to the coherent amplitude of 
$\alpha=0.97$. 
The fidelity between this output and 
$|\alpha\rangle + i |-\alpha\rangle$ is $F=0.993$. 
Note that this is the same as the state generated from 
a coherent state by strong Kerr nonlinear evolution \cite{Yurke86}. 
Figures~\ref{fig:WignerFunctions}(b)-(d) plot 
the Wigner functions with realistic detectors ($\eta=0.1$ and $\nu=10^{-7}$) 
for different $\{ c_+, c_- \}$. 
Even with such imperfect detectors, 
high fidelities ($> 0.95$) could be achieved.

\section{Amplification of the superposed states via 
conditional homodyne detection
\label{sec:amp}}
The fidelity between the state generated from our scheme 
and the ideal superposition of coherent states starts to 
decrease rapidly for $\alpha >1$. 
To realize surely macroscopic superposition or to apply these states 
to the coherent state superposition based quantum computation scheme 
\cite{Ralph03}, 
the superposed states are required to have larger amplitudes. 
One approach for the production of such a state is 
to introduce PNRDs in our scheme (Fig.~\ref{fig:SchemeOnOff}) 
to generate a superposition of $|\Psi_m\rangle$ and $|\Psi_{m+1}\rangle$ 
for large $m$. 
The other approach may be to apply the conditional amplification process 
proposed in \cite{Lund04}, which does not require PNRDs.  
It was shown that if one can prepare two inputs 
$|\alpha\rangle + e^{i \varphi} |-\alpha\rangle$ and 
$|\beta\rangle + e^{i \phi} |-\beta\rangle$, 
they can be conditionally transformed to 
the state $|\gamma\rangle + e^{i(\varphi+\phi)} |-\gamma\rangle$, 
where $\gamma = \sqrt{\alpha^2 + \beta^2}$, 
by using BSs, an auxiliary coherent state, 
and two on/off detectors. 
It could be used to amplify 
the initial state of $S(r)|1\rangle$, which well approximates 
$|C_-(\alpha)\rangle$ for $|\alpha|^2 \le 1$, 
and the scalability of the repetitive amplification process was 
discussed in detail in \cite{Jeong05}.

\begin{figure}
\begin{center}
\includegraphics[width=0.8\linewidth]{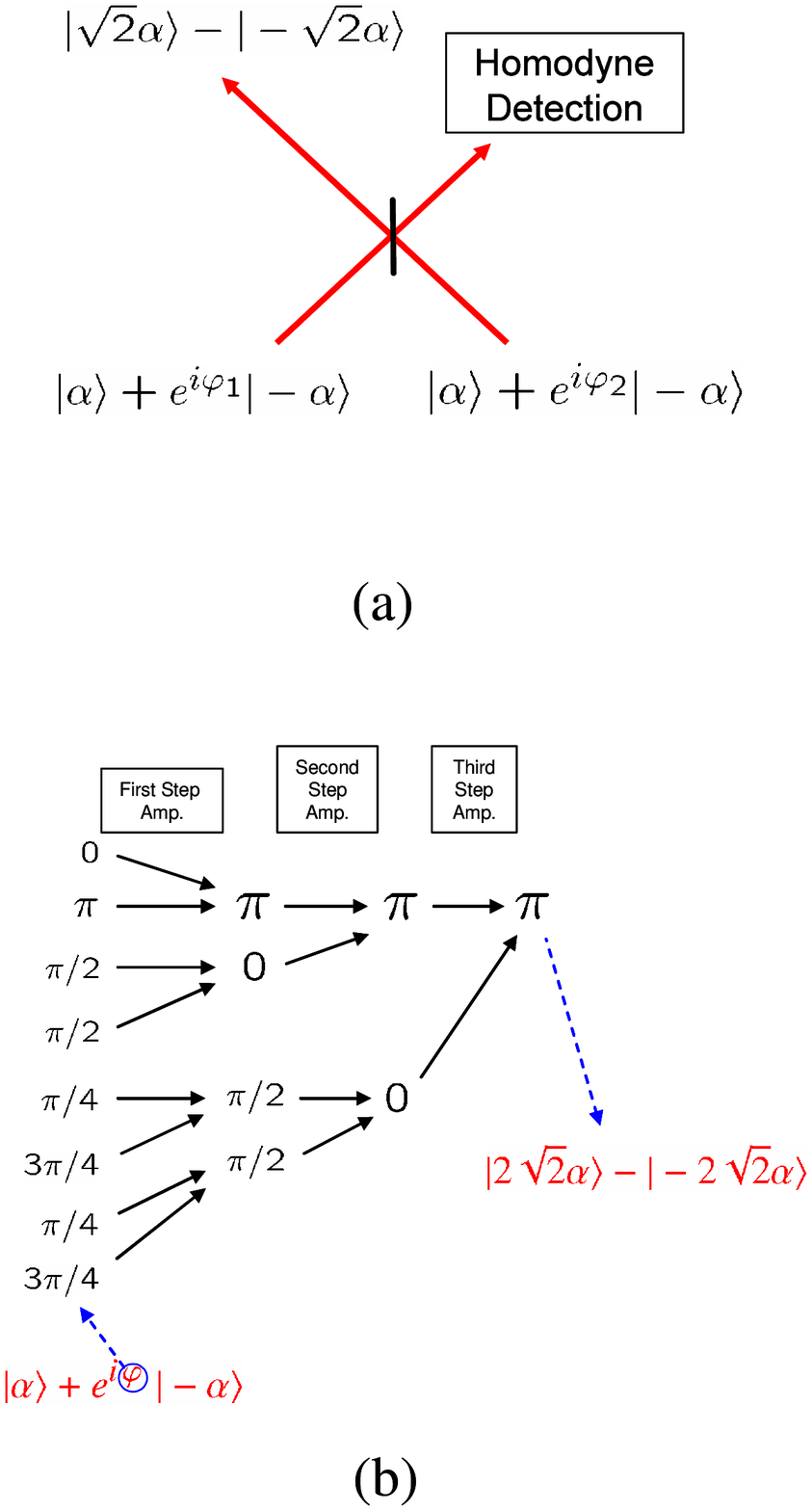}   %
\caption{\label{fig:CatAmpHD}
(a) A schematic of the conditional amplification of the coherent state 
superpositions with homodyne detection. (b) The three-step concatenation of 
the amplification process. 
The 8 initial states $|\alpha\rangle + e^{i\varphi_j} |-\alpha\rangle$ 
($j=1,...,8$) 
are prepared in certain phases $\varphi$ as illustrated in the left. 
}
\end{center}
\end{figure}

Application of our scenario to their scheme allows us 
to generate $|\gamma\rangle + e^{i \varphi}|-\gamma\rangle$ 
with large $\gamma$ and arbitrary $\varphi$. 
Moreover, we will show in this section that 
the two on/off detectors used 
in the scheme of Ref.~\cite{Lund04,Jeong05} 
can be simply replaced by a homodyne detector 
although the latter acts as a Gaussian operation. 
Note that it is not prohibited to transform non-Gaussian states 
to the other non-Gaussian states by only Gaussian operations. 
The conditional homodyne detection technique has been applied to 
purify the coherent state superpositions \cite{Suzuki06} 
or the squeezed states suffering non-Gaussian noises 
\cite{Heersink06}.

The schematic of the conditional amplification via homodyne detection 
is shown in Fig.~\ref{fig:CatAmpHD}(a). 
To explain how it works, let us see a simple example 
where we have $|C_+(\alpha)\rangle$ and $|C_-(\alpha)\rangle$ as 
two input states. 
These states are combined by a balanced BS as 
\begin{eqnarray}
\label{eq:interference_two_cats}
&& (|\alpha\rangle + |-\alpha\rangle)(|\alpha\rangle - |-\alpha\rangle)
\nonumber\\ &&
\stackrel{\rm BS}{\to}
(|\sqrt{2}\alpha\rangle - |-\sqrt{2}\alpha\rangle) |0\rangle
+ |0\rangle (|\sqrt{2}\alpha\rangle - |-\sqrt{2}\alpha\rangle) ,
\nonumber\\
\end{eqnarray}
where normalization factors are omitted for simplicity. 
Then one makes homodyne detection on one of the two modes. 
An ideal homodyne detection corresponds to a projection onto 
the quadrature eigenstate $|x\rangle$, and 
for the measurement outcome $x$, 
one obtains the conditional output state 
\begin{eqnarray}
\label{eq:x_conditioned_output}
&& \langle x|(|\sqrt{2}\alpha\rangle - |-\sqrt{2}\alpha\rangle) |0\rangle
+ \langle x|0\rangle (|\sqrt{2}\alpha\rangle - |-\sqrt{2}\alpha\rangle)
\nonumber\\
&& \propto 
\left( e^{-(x-2\alpha)^2/2} - e^{-(x+2\alpha)^2/2} \right) 
|0\rangle 
\nonumber\\ &&
\quad + e^{-x^2/2} 
\left( |\sqrt{2}\alpha\rangle - |-\sqrt{2}\alpha\rangle \right) . 
\end{eqnarray}
Here, conditioned on the outcome $x=0$, 
the first term vanishes and one obtains the amplified state 
$|\sqrt{2}\alpha\rangle - |-\sqrt{2}\alpha\rangle$. 
More generally, the condition for the two inputs 
$|\alpha\rangle + e^{i\varphi_{1,2}} |-\alpha\rangle$ 
to be amplified is $\varphi_1 + \varphi_2 = \pi$. 
The amplified state has the phase $\varphi_1 - \varphi_2$ 
which implies that one can choose arbitrary $\varphi$ 
at the output.

For further amplification, 
the process should be concatenated 
by carefully preparing the initial input states. 
Figure~\ref{fig:CatAmpHD}(b) is the schematic of the process to 
generate $|2\sqrt{2}\alpha\rangle - |-2\sqrt{2}\alpha\rangle$ 
by concatenating three amplification steps. 
The numbers represent the phase $\varphi$ of each state. 
The same rule can be applied in a straightforward way 
for the iterative generation of a superposition of large coherent states 
with arbitrary $\varphi$.

Homodyne detection is a well matured technique 
and very high quantum efficiency ($\eta \ge 0.99$) has been achieved 
with current technology. 
It simplifies the experimental complexity, 
enhances the practical success probability 
compared to a use of two imperfect on/off detectors, and thus 
will increase the total feasibility of the experimental demonstration.

\section{Conclusions\label{sec:conclusion}}
In this paper, a novel scheme for the conditional 
generation of a coherent state superposition with arbitrary 
ratio and phase has been proposed. 
The scheme uses a squeezed vacuum, beamsplitters, a coherent state 
ancilla, and two on/off detectors. 
We first showed that 
$c_+ |\alpha\rangle + c_- |-\alpha\rangle$ for arbitrary 
$\{ c_+,c_- \}$ can be approximated by 
an appropriate superposition of single- and two-photon subtracted 
squeezed vacua with very high fidelity ($F>0.99$).

Such a superposition of photon subtracted states is conditionally 
generated by using ideal photon number resolving detectors and 
a qubit ancilla $b_0|0\rangle + b_1|1\rangle$. 
We have shown that this ideal scheme is also realized 
by the more practical scheme
in which a displacement operation (coherent state ancilla) 
and on/off detectors are used. 
Even including realistic dark counts and low quantum efficiency 
(e.g. $\nu=10^{-7}$ and $\eta=0.1$), 
fidelities of more than $0.95$ could be achieved with 
a highly transmissive beamsplitter ($T=0.95$) and 
the squeezing of $r=0.3$ ($\sim 2.6$ dB), which 
corresponds to the amplitude of $\alpha = 0.95$ 
for the generated state. 
All these parameters are reasonably comparable with 
the recent experiments on single-photon subtraction 
\cite{Ourjoumtsev06,N-Nielsen06,Wakui06}.

We have also shown that our scheme is useful to simplify 
the setup of the conditional amplification of the coherent state 
superpositions originally proposed in \cite{Lund04,Jeong05}. 
Our simplified version is quite feasible to demonstrate 
with current experimental techniques. 
Our scheme would also be useful to save 
the amount of required resources in other quantum information 
applications, including superposed coherent state based quantum computing 
\cite{Ralph03,Lund05}.


\begin{references}
\bibitem{Comment1}
Note that the definition of $|\Psi_m^{(-)}\rangle$ in our paper 
is slightly different from the Dakna's original one \cite{Dakna97} 
by a factor of $(-1)^m$. 
Here we also refer the photon number expansion of $|\Psi_m\rangle$ as 
\cite{Dakna97}
$$
|\Psi_m\rangle = \frac{1}{\sqrt{\mathcal{N}_m}} 
\sum_{n=0}^\infty a_{m,n} |n\rangle ,
$$
where 
\begin{eqnarray}
\label{eq:C}
a_{m,n} & = & \frac{(m+n)!}{\Gamma[(m+n)/2 + 1] \sqrt{n!}} ,
\nonumber\\ && 
\times \frac{1+(-1)^{m+n}}{2} 
\left( \frac{\lambda T}{2} \right)^{(m+n)/2} , 
\nonumber
\end{eqnarray}
and 
\begin{eqnarray}
\label{eq:N_m}
\mathcal{N}_m & = & \frac{1}{\sqrt{1-\lambda^2 T^2}} 
\left[ \frac{\lambda^2 T^2}{1-\lambda^2 T^2} \right]^m 
\nonumber\\ && 
\times \sum_{k=0}^{[m/2]} \frac{(m!)^2}{(m-2k)!(k!)^2(2\lambda T)^{2k}} .
\nonumber
\end{eqnarray}
\end{references}
\end{document}